\documentclass[twocolumn,superscriptaddress,aps,prl]{revtex4-1}

\usepackage{graphicx}
\usepackage{color}
\usepackage[colorlinks=true,
            linkcolor=blue,
            urlcolor=blue,
            citecolor=blue,
            hyperfootnotes=true]{hyperref}

\newcommand{\teq}{{\,=\,}}
            
\begin{document}
\title{Stable charge density wave phase in a 1T-TiSe$_2$ monolayer}

\author{Bahadur Singh}
\affiliation{Centre for Advanced 2D Materials and Graphene Research Centre, National University of Singapore, Singapore 117546}
\affiliation{Department of Physics, National University of Singapore, Singapore 117542} 

\author{Chuang-Han Hsu}
\affiliation{Centre for Advanced 2D Materials and Graphene Research Centre, National University of Singapore, Singapore 117546}  
\affiliation{Department of Physics, National University of Singapore, Singapore 117542} 

\author{Wei-Feng Tsai}
\affiliation{Centre for Advanced 2D Materials and Graphene Research Centre, National University of Singapore, Singapore 117546}  
\affiliation{Department of Physics, National University of Singapore, Singapore 117542}  
\affiliation{School of Physics, Sun Yat-sen University, Guangzhou, China 510275}

\author{Vitor M. Pereira}
\thanks{Corresponding author: vpereira@nus.edu.sg}
\affiliation{Centre for Advanced 2D Materials and Graphene Research Centre, National University of Singapore, Singapore 117546}  
\affiliation{Department of Physics, National University of Singapore, Singapore 117542} 

\author{Hsin Lin}
\affiliation{Centre for Advanced 2D Materials and Graphene Research Centre, National University of Singapore, Singapore 117546}  
\affiliation{Department of Physics, National University of Singapore, Singapore 117542}

\begin{abstract}
Charge density wave (CDW) phases are symmetry-reduced states of matter in which a periodic modulation of the electronic charge frequently leads to drastic changes of the electronic spectrum, including the emergence of energy gaps. We analyze the CDW state in a 1T-TiSe$_2$ monolayer within a density functional theory framework and show that, similarly to its bulk counterpart, the monolayer is unstable towards a commensurate $2{\times}2$ periodic lattice distortion (PLD) and CDW at low temperatures. Analysis of the electron and phonon spectrum establishes the PLD as the stable $T=0$\,K configuration with a narrow bandgap, whereas the undistorted and semi-metalic state is stable only above a threshold temperature. The lattice distortions as well as the unfolded and reconstructed band structure in the CDW phase agree well with experimental results. We also address evidence in our results for the role of electron-electron interactions in the CDW instability of 1T-TiSe$_2$ monolayers.
\end{abstract}

\maketitle

Recent years have witnessed remarkable progress in the study of two dimensional (2D) materials owing to their diverse properties and potential applications \cite{2D_Nov2005,2DO_Butler2013,2DH_Nov2016}. Due to the reduced dimensionality, their physics can differ significantly from the bulk counterparts, while providing greater flexibility for tuning their electronic properties through changing the number of layers, chemical composition or by their integration in heterostructures \cite{2DH_Nov2016,TMDC_hetrostr}. The 2D thin films of transition metal dichalcogenides (TMDs) with chemical formula MX$_2$ (where M is a transition metal and X is a chalcogen) are particularly appealing because they offer a wealth of electronic properties ranging from insulating to semiconducting to metallic or semimetallic, depending on the choice of transition metal or chalcogen \cite{2D_Nov2005,2DO_Butler2013,2DH_Nov2016,TMDC_hetrostr,MoS2_NewM, MoS2_DIBansil,TMDC_CDWReview,TaS2_Yu2015,TMDC_CDWThin2015}. Their different electronic behavior generally arises from the partially filled d-bands of the transition metal ion. In addition, some of the layered TMDs are found to exhibit generic instabilities towards the symmetry-lowering charge density wave (CDW) state and superconductivity and, therefore, are ideal platforms to investigate in a controlled manner the interplay between these phases \cite{2DH_Nov2016,TaS2_Yu2015,FermiSurfNest,TMDC_CDWThin2015,TiSe2M_superCE2016, TiSe2M_chrialSup,2D_Nov2005,2DO_Butler2013,2DH_Nov2016,MoS2_Wang2012, 
MoS2_ValleyDirac,TMDC_FET}. 

1T-TiSe$_2$ (henceforth TiSe$_2$, for simplicity) is among the most studied TMDs due to its simple commensurate $2{\times}2{\times}2$ CDW state below $T_c\simeq200$ K in the bulk \cite{TiSe2B_expN,TiSe2B_expARPES,TiSe2B_expARPES_EHC,TiSe2B_exp_Exct, TiSe2B_JahnTell,TiSe2B_ThVCD,TiSe2B_ThSm,TiSe2B_Th,TiSe2B_ThMech,Cazzaniga2012,Olevano2014}. It has been established that the CDW order in the bulk is weakened by either Cu intercalation or pressure, and that a dome-like superconducting phase appears near the point of CDW suppression in either phase diagram, indicating a tight interplay between the CDW order and superconductivity in bulk TiSe$_2$ \cite{TiSe2B_supCu,TiSe2B_supPress}. The dominant underlying mechanism for the CDW transition in this material has been a subject of intense theoretical and experimental study for more than three decades, and the question remains unsettled. Several experimental studies have suggested that either an excitonic interaction and/or band Jahn-Teller effect is responsible for the CDW instability \cite{TiSe2B_expARPES_EHC,TiSe2B_exp_Exct,TiSe2B_JahnTell,TiSe2B_ThSm,TiSe2B_Th,TiSe2B_ThMech,Cazzaniga2012,Olevano2014}. The difficulties to reach a consensus possibly arise because of the 3D nature of the CDW order which makes it difficult to identify the exact gap locations in the 3D Brillouin zone (BZ). In contrast, this problem seems to be more tractable in the case of the TiSe$_2$ monolayer because of its intrinsically 2D band structure that facilitates the experimental analysis of spectral weight transfer and gap opening. Recently, thin films of TiSe$_2$ have been fabricated and experimentally found to exhibit a $2{\times}2$ CDW ordering below a critical temperature that can be controlled in few-layer samples by changing the film thickness and/or field effect \cite{TiSe2M_raman,TiSe2M_CDWSTM,TiSe2M_CDWNC,TiSe2M_CDWNano,TiSe2M_superCE2016,TiSe2M_chrialSup}. Furthermore, the superconducting dome remains in the thin films, and field effect doping can be used to reach it and tune the superconducting transition temperature \cite{TiSe2M_superCE2016}. These studies suggest that monolayer TiSe$_2$ will not only help explaining the CDW mechanism in the bulk, but constitutes an interesting system on its own as a prototypical 2D material to investigate the interplay between these collective phenomena. 

In this paper, we describe systematic {\it ab-initio} electronic structure calculations to investigate the periodic lattice distortion (PLD) and CDW ordering, as well as their underlying mechanism in the TiSe$_2$ monolayer. The normal $1{\times} 1$ phase is seen to be unstable at low temperatures due to the softening of a zone-boundary phonon mode at the $M$-point. This phonon's frequency depends strongly on the electronic smearing parameter (electronic temperature), indicating a structural phase transition with temperature. We find that the $2{\times}2$ superstructure is the ground state at $T=0$ K, whereas the normal $1{\times}1$ structure is stabilized only at higher temperatures. The unfolded band structure shows an energy gap coinciding with the Fermi level $E_F$, as well as clear backfolded bands at the $M$-point, in excellent agreement with experimental energy dispersions obtained from angle-resolved photoemission spectroscopy (ARPES). Our results clearly demonstrate that the CDW formation in the TiSe$_2$ monolayer is intimately associated with a robust structural phase transition that reduces the lattice symmetry at low temperatures. 

\begin{figure}[tb] 
\includegraphics[width=0.5\textwidth]{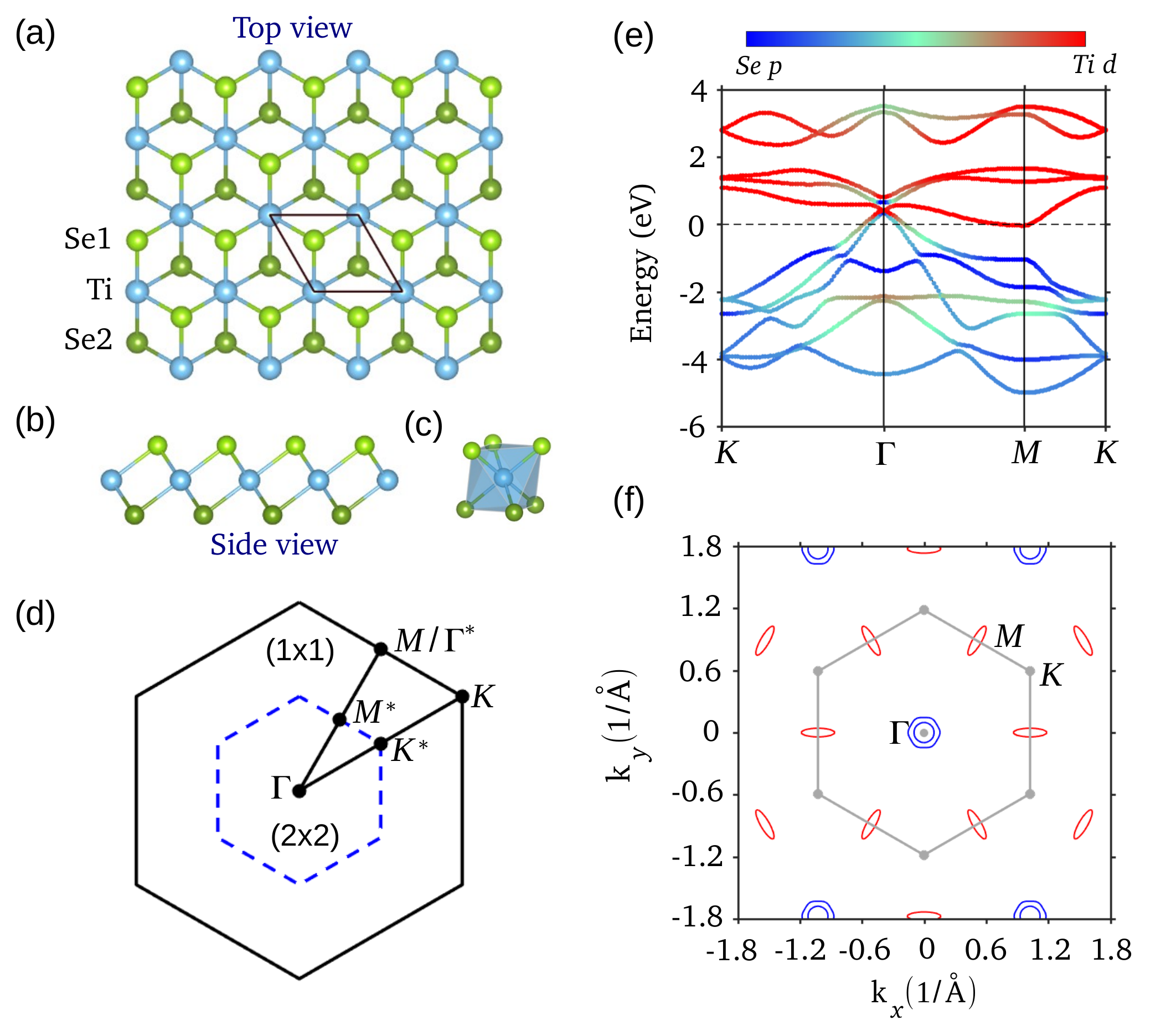} 
\caption {(a) Top and (b) side view of a TiSe$_2$ monolayer structure. Blue and green balls denote Ti and Se atoms, respectively. Se1 (light green) and Se2 (dark green) represent Se atoms in the top and bottom sublayers, respectively. (c) Local octahedral structure of TiSe$_2$ monolayer. Ti atoms lie at the center of the octahedral coordination unit. (d) First Brillouin zones (BZs) of the monolayer in the normal phase (black solid lines) and the CDW phase (blue dashed lines) with three high symmetry points $\Gamma$, $M$, and $K$. The high symmetry points of the reduced BZ are marked with a star ($\ast$). (e) Band structure of monolayer TiSe$_2$ in the $1 {\times} 1$ normal phase. Ti $d$ and Se $p$ orbital characters are shown by red and blue colors, respectively. (f) The corresponding constant energy contours at $E_F$. Blue and red color represent hole and electron pockets, respectively.} \label{fig:NormalBS}
\end{figure}

\section{Methodology}

Electronic structure calculations were performed with the projector augmented wave method \cite{vasp,paw} within the density functional theory (DFT) \cite{kohan_dft} framework, using the VASP code \cite{vasp}. According to our calculations, results obtained with the GGA (generalized gradient approximation) \cite{pbe} functional for the exchange-correlation (XC) effects agree better with the experimental data (namely, the magnitude of the atomic displacements and the energy spectrum reconstruction, to be discussed in detail below). Therefore, we report here the results obtained with the GGA except where explicitly stated otherwise. The spin-orbit coupling was included self-consistently. Lattice parameters and ionic positions were optimized until the residual forces on each ion were less than  $1.0\times 10^{-3}$\,eV/\r{A}. We obtained an optimized in-plane lattice constant of $a \teq 3.536$ \r{A}, in close agreement with the experimental value ($a \teq 3.538$\,\r{A}) \cite{TiSe2M_CDWSTM,TiSe2M_CDWNC} and earlier theoretical results for bulk \cite{TiSe2B_Th}. This good agreement leads to negligible spurious external pressure in the calculations which, as noted by Olevano \emph{et al.} \cite{Olevano2014}, is crucial for a reliable prediction of the CDW instability based on phonon calculations. In view of this agreement, all calculations were performed using a slab model with the experimental lattice constant and fully optimized ionic positions. We used a vacuum region of 12\,\r{A} to avoid interaction between the periodically repeated slabs, and a plane-wave cutoff energy of 380\,eV throughout the  calculations.

The phonon dispersion curves were computed using density functional perturbation theory (DFPT) \cite{phonon_DFPT} as implemented in the PHONOPY code \cite{phonopy} with a $2{\,\times\,}2$ supercell. Convergence tests with 
respect to $k$-point sampling within the normal and the superstructure BZ were carried out for both electronic and vibrational properties. The convergence was reached with a $\Gamma$-centered $16{\,\times\,}16{\,\times\,}1$ $k$-mesh which is that ultimately used in the calculations reported below.  All the ground state calculations were done with a small smearing parameter ($\sigma \teq 0.001$\,eV) which was well converged with respect to different smearing functions \cite{vasp,Sm_FD,Sm_gauss,Sm_MethPax}. Unfolding of the band structure was done using a home-built code based on Ref. ~\cite{unfold_popeZung}. As this method has been thoroughly 
discussed in earlier publications \cite{unfold_popeZung,unfold_Chicheng,unfold_paulo1,unfold_paulo2}, for the sake of brevity and to avoid repetition, we do not discuss its details here. 

\section{The normal phase}

The TiSe$_2$ monolayer has an hexagonal Bravais lattice with the space group $D^3_{3d}$ ($P\bar{3}m1,164$) \cite{TiSe2M_CDWSTM,TiSe2M_CDWNC}. It consists of three sublayers stacked in the order Se-Ti-Se within a single unit cell, as shown in Figs.~\ref{fig:NormalBS}(a)-\ref{fig:NormalBS}(c). The two Se sublayers are strongly bonded with the Ti plane in the middle (bond length $\simeq$ 2.56 \r{A}) and Ti has an octahedral prismatic coordination as illustrated in Fig.\ \ref{fig:NormalBS}(c). The first BZ is hexagonal with three high symmetry points $\Gamma$, $M$, and $K$, as shown in Fig.\ \ref{fig:NormalBS}(d). This BZ can be obtained by projecting the 3D bulk BZ onto the (001) surface and, therefore, the points corresponding to the CDW wavevector in the bulk ($L$ and $M$ in the 3D BZ) map into $M$ points of the 2D BZ. The PLD doubles the original lattice periodicity forming a $2{\times}2$ superstructure and, hence, reduces the 2D BZ, which is shown by broken blue lines in Fig.\ \ref{fig:NormalBS}(d). Due to the BZ folding, the original $M$ points of the $1{\times}1$ BZ are mapped into the zone center ($\Gamma^*$) of the $2{\times}2$ BZ. 

\begin{figure}[tb] 
\includegraphics[width=0.5\textwidth]{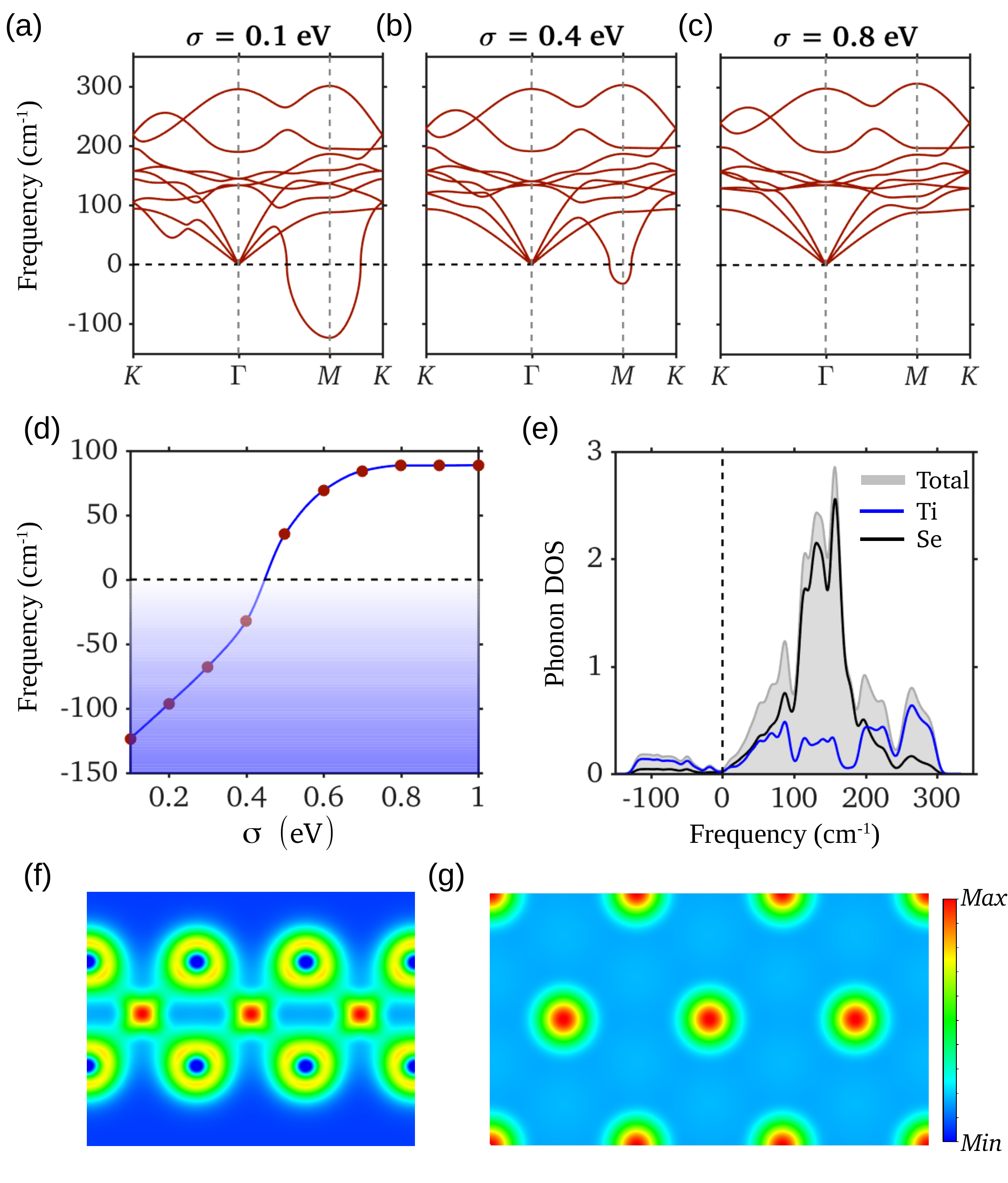} 
\caption{(a)-(c) Phonon dispersion of the $1{\times}1$ TiSe$_2$ monolayer with different electronic smearing parameter $\sigma$. The frequency of the soft phonon mode at $M$-point changes from imaginary (represented with a negative value) to a positive value with increasing $\sigma$. (d) Evolution of the lowest phonon mode frequency at the $M$-point as a function of $\sigma$. The strong dependence of this phonon's frequency on $\sigma$ and its quenching signals a CDW phase transition. (e) Total and partial phonon density of states for $\sigma=0.1$ eV. The imaginary (negative) phonon frequencies arise mainly from the Ti atoms. (f) Charge density distribution in the Ti-Se plane containing Ti-Se bonds and (g) in the Ti plane of an undistorted $2{\times}2$ superstructure of TiSe$_2$ monolayer.} 
\label{fig:NormalPhonon}
\end{figure}

Beginning with a  monolayer in the normal phase, a simplistic ionic insulator model suggests a strong tendency for electron transfer from the Ti $3d$ and $4s$ orbitals to the $4p$ states of Se. Valence and conduction bands would then arise from Se $4p$ and Ti $3d$ states, respectively. However, in reality, the crystalline environment and spatial extent of the $d$ orbitals increase their bandwidth such that conduction and valence bands overlap. Figure \ref{fig:NormalBS}(e) shows that, near the Fermi level, the Se $4p$-derived valence bands are centered at the $\Gamma$-point whereas the Ti $3d$-derived conduction bands lie at the $M$-point. These overlap in energy and make TiSe$_2$ a semimetal in the normal phase. The Fermi contours reveal a pair of hole pockets at the $\Gamma$-point whereas a single elongated elliptical electron pocket forms at the $M$-point, as shown in Fig.~\ref{fig:NormalBS}(f). In order to equivalently quantify a band overlap or band gap, we define the indirect gap as the difference between the minimum of the conduction band at the $M$-point and the maximum of the valence band at $\Gamma$. A negative value of the gap therefore represents a semimetal with indirect band overlap, while an insulator will have a positive gap. We find an indirect band gap of $-0.446$ eV using GGA, and a smaller value of $-0.242$ eV with an HSE functional. Although these values are different from some experimentally reported values of $\sim+$0.098 eV \cite{TiSe2M_CDWNC,TiSe2M_CDWNano}, they agree well with earlier bandstructure calculations \cite{TiSe2M_CDWNC}. The difference with respect to the experimental data may arise because our DFT calculations assume $T = 0$ K whereas the $1{\times}1$ normal structure, in reality, exists only at higher temperatures.   

The phonon dispersion of the normal phase is shown in Fig.~\ref{fig:NormalPhonon} for different values of the electronic smearing parameter $\sigma$ using a Methfessel-Paxton smearing \cite{Sm_MethPax}. This parameter determines the smearing width and is normally used as a technical tool to accelerate convergence in DFT calculations. However, when used with the Fermi-Dirac distribution, it mimics the electronic temperature and thus 
describes the occupation probability of the electronic states \cite{TiSe2B_ThSm,Sm_FD,Sm_MethPax}. By varying $\sigma$, we can qualitatively estimate changes expected to occur in the phonon spectrum with increasing 
temperature, thereby monitoring the structural stability at different temperatures. In Fig.\ \ref{fig:NormalPhonon}(a) we show the phonon spectrum at a small smearing parameter. It is clear that the system is dynamically unstable with a Kohn-type \cite{kohn_anoM} soft mode at the $M$-point. The partial phonon density of states is shown in Fig.\ \ref{fig:NormalPhonon}(e) and demonstrates the imaginary frequencies are inherent to the Ti atoms, suggesting that the CDW is associated with the Ti sublayers. The structural instability at the $M$-point is consistent with the $2{\times}2$ commensurate PLD observed in experiments at low temperature. As we increase $\sigma$, the range over which the soft mode has imaginary frequency is reduced, and finally disappears for $\sigma \sim 0.4-0.5 $ eV [see Fig.\ \ref{fig:NormalPhonon}(a)-\ref{fig:NormalPhonon}(d)]. This dependence of the soft mode frequency on $\sigma$ indicates a structural phase transition with temperature and confirms that the $1{\times}1$ structure is stable only above a threshold temperature \cite{footnote1,footnote2}. 

\begin{figure*}[t] 
\centering
\includegraphics[width=0.8\textwidth]{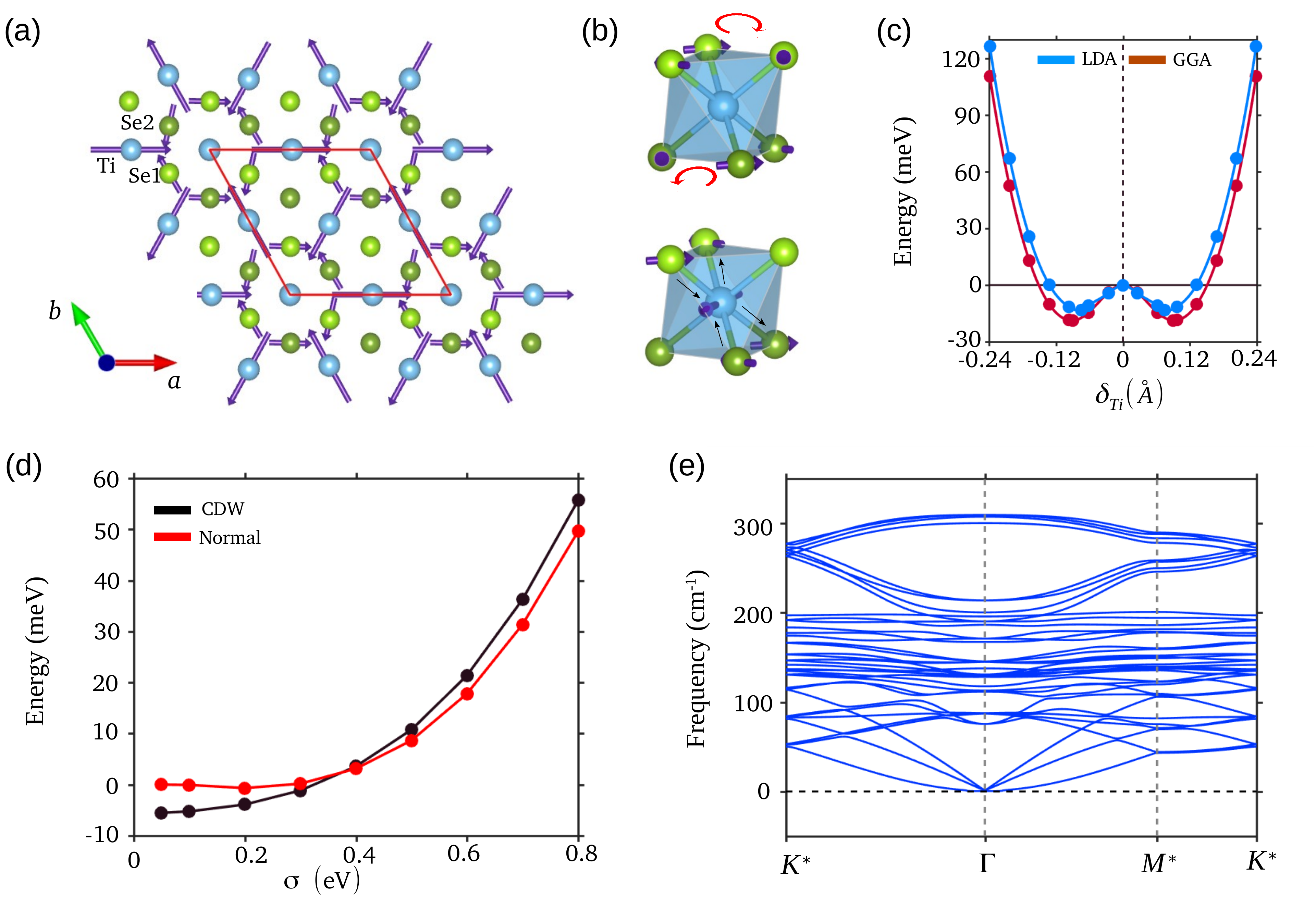} 
\caption{(a) Illustration of the atomic displacements in the CDW phase. The length of each arrow is proportional to the magnitude of the atomic displacement. (b) The two types of local octahedral structure in the CDW state. 
Red arrows in the first depict the movement of Se atoms, while the inward and outward black arrows in the second illustrate the contraction and expansion, respectively, of the Ti-Se bonds. (c) Evolution of the total energy of the $2{\,\times\,}2$ superstructure computed using LDA and GGA as a function of CDW distortion $\delta_{Ti}$. Energies are given relative to the normal phase energy ($\delta_{Ti} \teq 0$). (d) Energy per chemical unit of the normal phase (red) and distorted phase (black) as a function of smearing parameter $\sigma$. Energies are given relative to the normal phase energy at $\sigma\teq 0.05$ eV. (e) Phonon dispersion curve of the distorted phase ($2{\,\times\,}2$ superstructure) in the supercell BZ.} 
\label{fig:cdwph}
\end{figure*}

Phonons and electrons are inseparably intertwined in a crystal which prevents, in principle, the simplistic assignment of the CDW/PLD tendency to an instability of the electronic or phonon subsystems (through electron-phonon 
coupling), independently. In a simplistic description of this ``hierarchy'', a transition driven entirely by electronic interactions, the PLD would be regarded ``secondary'',  as a readjustment of the ions to a modified Born-Oppenheimer potential \cite{TMDC_CDWReview,FermiSurfNest,TiSe2B_ThMech}, or it might not even occur at all, as seems to be the case, for example, in the layered purple bronze K$_{0.9}$Mo$_6$O$_{17}$ \cite{Mou2016,Su2016}. The non-uniform charge density would therefore emerge regardless of whether or not the ions are clamped at their high symmetry positions \cite{FermiSurfNest,TiSe2B_ThMech}. Figures \ref{fig:NormalPhonon}(f)-\ref{fig:NormalPhonon}(g) show the charge density distribution in various crystal planes in an undistorted $2{\times}2$ superstructure of TiSe$_2$ monolayer. They show no appreciable charge redistribution in the presence of a doubled lattice periodicity, similarly to the previously studied case of the bulk material \cite{TiSe2B_ThMech}. This, however, must be interpreted with care, as a consistency check, and not as confirmation that the role of electronic interactions is secondary. Such conclusion would be rather primitive because, on the one hand, systems where CDW arise only from electronic interactions are usually strongly correlated and, on the other, the GGA XC functional cannot capture correlations at that level. In fact, in bulk TiSe$_2$, the inclusion of many-body corrections at the level of the $GW$ approximation has been shown to capture the spectral reconstruction and gapped state seen experimentally even without any deformation (i.e. in a $1{\,\times\,}1$ cell calculation) \cite{Cazzaniga2012}, which underscores that electronic correlations are indeed of crucial importance to describe the electronic state through the CDW transition. The charge distributions seen in Fig.~\ref{fig:NormalPhonon}(g)] are spherically symmetric near the Ti atoms and nearly constant in the interstitial region. This is a reminiscent feature of the metallic-type ionic environment and suggests that Ti atoms are more likely to displace in their hexagonal plane to find an energy minimum.

\section{The distorted phase}

We now investigate the $2\times2$ superstructure to find the equilibrium configuration in the distorted phase. We have allowed all the ions' positions to relax using both the LDA and GGA functionals. Figures \ref{fig:cdwph}(a)-\ref{fig:cdwph}(b) illustrate the atomic movement in the fully relaxed $2\times2$ superstructure. Both the GGA and LDA predict the CDW instability, with an energy reduction of $\sim$4.7 meV and $\sim$3.7 meV per 
chemical unit, respectively. Interestingly, despite unconstrained, our results show that all atoms move only in their respective atomic planes without any out-of-plane distortion. The in-plane atomic displacements give rise to two different local octahedral structures in the $2\times2$ superstructure: in one octahedron, Ti atoms remain the center of the coordination unit while top (bottom) Se atoms are displaced clockwise (counterclockwise) in a circular fashion without affecting the original Ti-Se bond length, as depicted in the Figure. In contrast, Ti atoms in the second octahedron are displaced off-center giving rise to a distorted octahedron with three different Ti-Se bond lengths, as shown in the lower part of Fig.\ \ref{fig:cdwph}(b). Even though our calculations with either the GGA or LDA functional yield similar atomic displacement patterns, the magnitude of the atomic displacements depends strongly on the functional used. The calculated atomic displacements $\delta_{Ti}$ ($\delta_{Se}$) for Ti (Se) are 0.090 (0.029) \r{A} and 0.076 (0.016) \r{A} with the GGA and LDA functional, respectively. The atomic displacement ratio $\delta_{Ti}/\delta_{Se}$ with the GGA functional is ${\,\approx\,}3.10$, which agrees well with experiments \cite{TiSe2B_expN,TiSe2M_CDWNC}.

\begin{figure}[t] 
\includegraphics[width=0.5\textwidth]{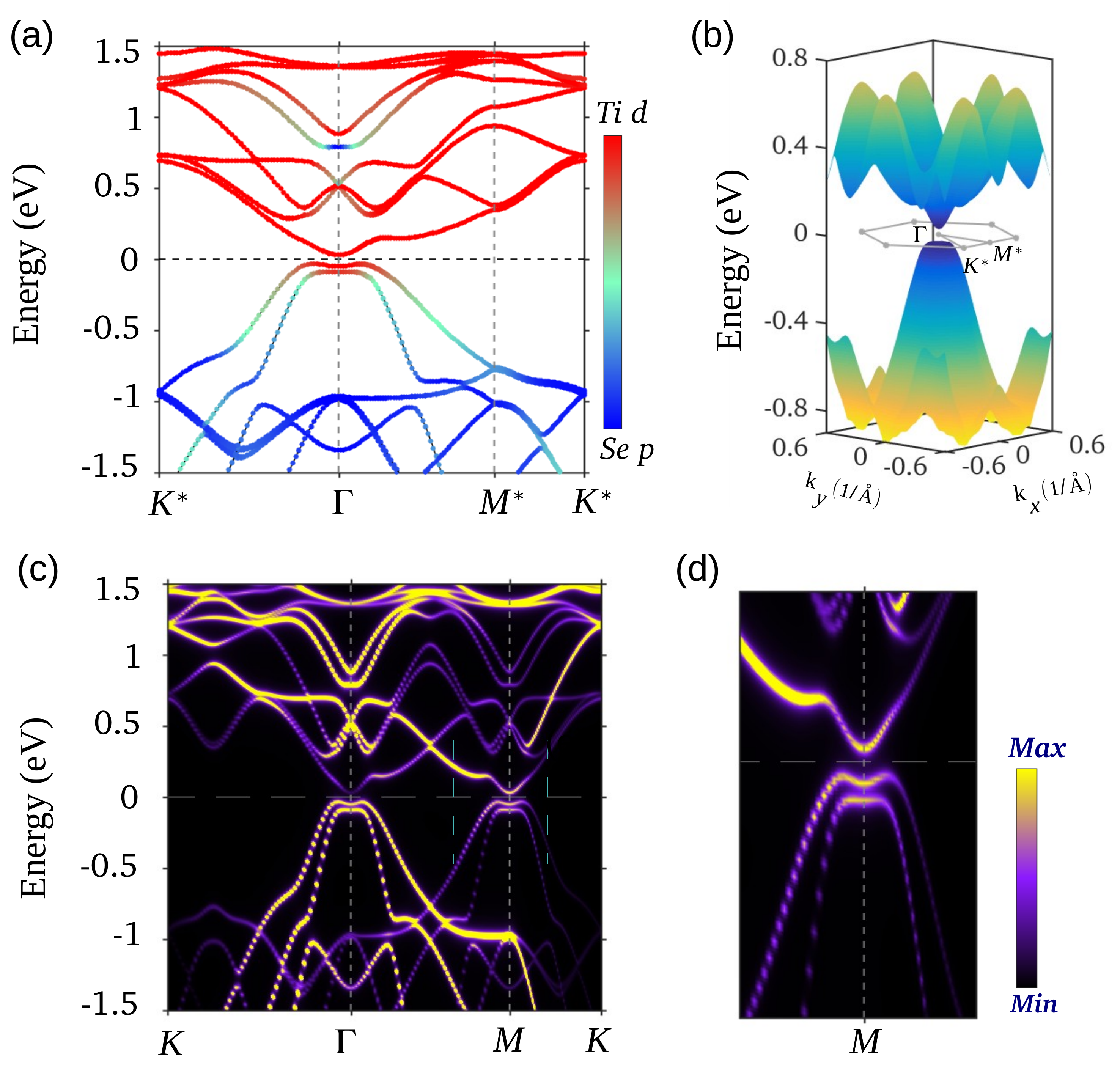} 
\caption{(a) Electronic band structure and (b) valence and conduction bands of the TiSe$_2$ monolayer in the distorted phase (relaxed $2\times 2$ superstructure), represented in the supercell BZ. (c) The corresponding 
unfolded band structure. (d) Close-up of the area highlighted by the dashed rectangle in (c).}
\label{fig:cdwBS}
\end{figure}

Fig.~\ref{fig:cdwph}(d) presents a complementary analysis of the relative stability of the distorted and undistorted configurations in terms of the energies per chemical unit of each phase obtained for different smearing 
parameters $\sigma$. The distorted phase has lower energy than the normal phase for small $\sigma$, and is thus more stable at lower temperature. As we increase $\sigma$, the energy of the distorted phase increases and surpasses that of the normal phase at $\sigma{\,\sim\,}0.4-0.5$\,eV. This behavior is in accord with the normal phase phonon analysis presented above [cf. Fig. \ref{fig:NormalPhonon}] and further confirms the stability of the $1{\,\times\,}1$ normal phase structure above a threshold temperature.

The phonon spectrum of the fully relaxed $2\times2$ superstructure is shown in Fig.\ \ref{fig:cdwph}(e). The absence of imaginary frequencies in the whole 2D BZ reflects the dynamical stability of this configuration at $T=0$ K, and indicates that it is the ground state structure of the TiSe$_2$ monolayer. This is further supported by the local energy landscape of the CDW phase that we analyzed by changing the atomic distortion $\delta_{Ti}$ manually and computing the changes in total energy; the results are shown in Fig.\ \ref{fig:cdwBS}(c). It should be noted that the Se atoms were fully relaxed for each manually set $\delta_{Ti}$. The energy is minimal at a finite value of $\delta_{Ti}$ (in the plane), and a spatial reversal of the distortion yields a degenerate configuration. The system can then freeze in either configuration at low temperature. At higher temperature, however, thermal effects allow the system to fluctuate between configurations giving rise to an ``average'' $1\times1$ structure characteristic of the normal state.

Figure \ref{fig:cdwBS}(a) displays the electronic band structure of the relaxed $2\times2$ superlattice, and shows the emergence of a full band gap in the BZ [see Fig. \ref{fig:cdwBS}(b)] at the Fermi level. The orbital character of the Bloch states is represented by the color map superimposed on each curve. It is clear that the coupling between the predominantly Ti-derived conduction band orbitals at $M$ and the predominantly Se-derived valence orbitals at $\Gamma$ lifts the band overlap that is present in the normal state and lowers (raises) the energy of the filled (empty) states in the vicinity of $E_F$ that becomes gapped.

In order to facilitate direct comparison with experimental dispersions \cite{TiSe2M_CDWNC,TiSe2M_CDWNano}, in Figs.\ \ref{fig:cdwBS}(c)-\ref{fig:cdwBS}(d) we unfolded the superlattice band structure to the original $1\times1$ BZ. The most significant feature in this representation is the clear presence of back-folded bands at the $M$-point which, despite their smaller spectral weight, provide a prominent signature of the new periodic potential in the CDW phase. The CDW phase in TiSe$_2$ has been recently investigated experimentally by ARPES, whose spectra reveal the formation of a $2{\times}2$ superlattice with a band gap of $\simeq\,$ 153\,meV at 10\,K and two back-folded bands at the $M$-point \cite{TiSe2M_CDWNC,TiSe2M_CDWNano}. The spectral weight of the back-folded bands at the {\it M}-point is smaller than that of the bands at $\Gamma$ \cite{TiSe2M_CDWNC}. These experimental results are well captured by our first-principles results. The insulating electronic state with a band gap of 82\,meV (325\,meV) with GGA (HSE) and the location and intensity of the two back-folded bands at the $M$-point are in reasonable agreement with those experiments.

Finally, we highlight the fact that, in addition to obtaining magnitudes of the lattice distortion and electronic gap which are accurate in comparison with experiments, we obtain also the non-trivial restructuring of the bands around $E_F$ that has been analyzed in detail on the basis of ARPES spectra in bulk TiSe$_2$ \cite{TiSe2B_expARPES_EHC,TiSe2B_exp_Exct}. This is best seen in the close-up of Fig.\ \ref{fig:cdwBS}(d) that shows the lowest conduction band around $\it{M}$ remaining parabolic, whereas the valence band acquires a Mexican-hat type dispersion. Combined with finite temperature broadening, the latter causes a flattening of the top of the valence band, an effect that has been seen clearly by ARPES in bulk samples \cite{TiSe2B_expARPES_EHC,TiSe2B_exp_Exct}. The qualitative significance of this reconstruction within a DFT calculation was first highlighted by Cazzaniga \emph{et al.} in studies of the distorted phase in the bulk \cite{Cazzaniga2012}. In that case, $GW$ corrections on top of the LDA are seen to capture the experimental Mexican-hat reconstruction even in an undistorted $1{\,\times\,}1$ cell. It is very interesting that, contrary to the bulk case, the monolayer shows this reconstruction already at the GGA level, Fig.~\ref{fig:cdwBS}(c), without needing to add many-body corrections beyond the XC functional. As argued in detail in Ref. \cite{Cazzaniga2012}, this fact strongly supports the built-in tendency of this electronic system towards an excitonic-insulator state, which predicts precisely such type of spectral reconstruction \cite{Kohn_EI}. It is noteworthy that such physics is captured already at the GGA level in the monolayer which, providing a more rudimentary account of the electronic interactions than the $GW$ approximation, perhaps suggests a stronger tendency for the excitonic instability in the monolayer in comparison with its bulk counterpart.

\begin{figure}[t] 
\includegraphics[width=0.5\textwidth]{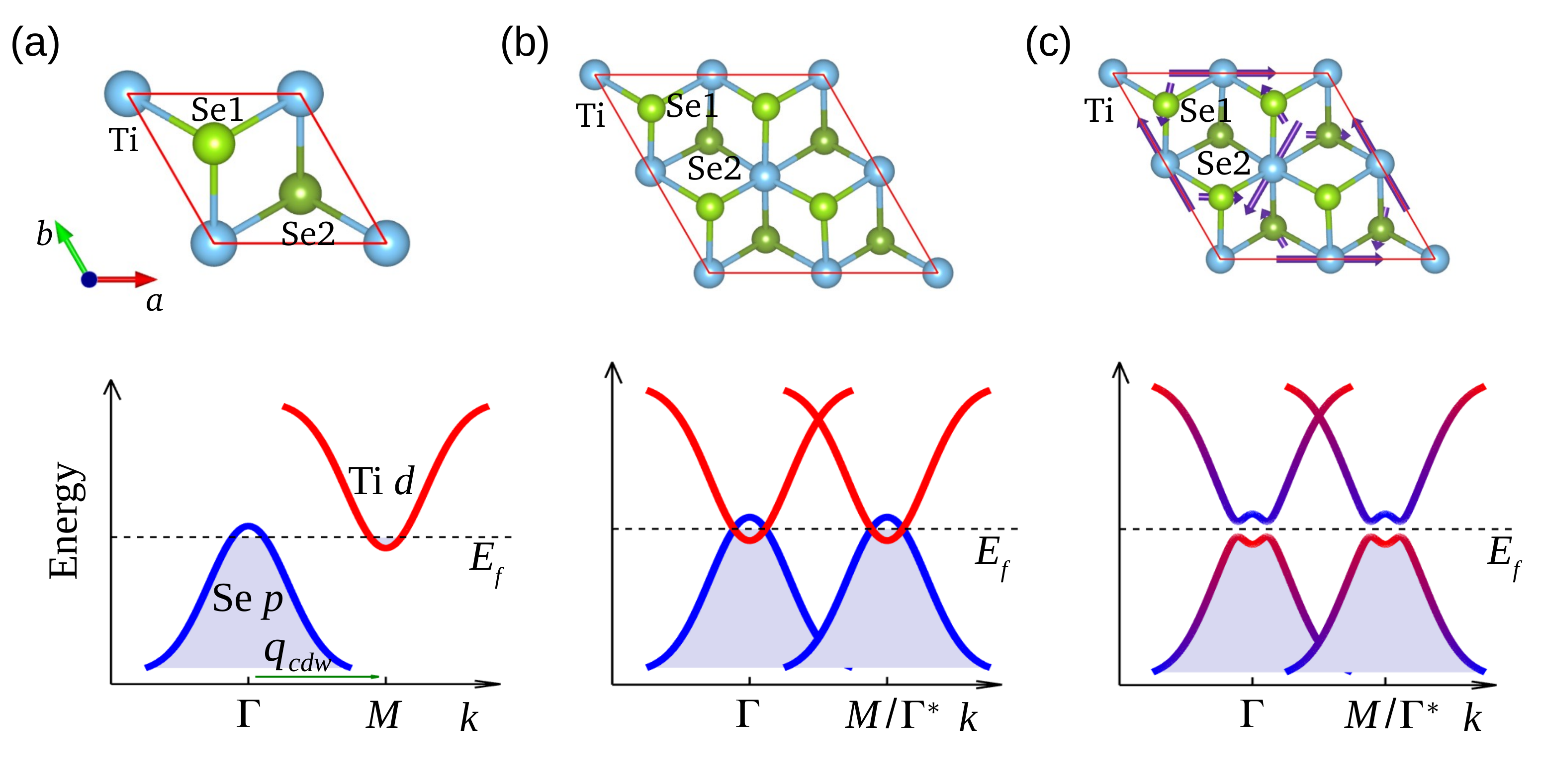} 
\caption{Schematic evolution of the CDW phase in a TiSe$_2$ monolayer (top row) along with a schematic band structure (bottom row). (a) In the normal phase, Se $4p$ derived hole pockets and Ti $3d$ derived electron pockets exist near the $\Gamma$ and $M$ point, respectively. (b) In a clamped-ion $2\times2$ superstructure, the electron and hole pockets are simply replicated (folded) at the $\Gamma$ and $M$-points without gapping the spectrum. (c) A finite lattice distortion develops when the ions are fully relaxed and a gap simultaneously opens at $E_F$. (a) Undistorted lattice with an elementary unit cell, (b) undistorted lattice with a $2{\,\times\,}2$ unit cell, and (c) periodically distorted lattice.}  
\label{fig:schBS}
\end{figure}

\section{Discussion and concluding remarks}

The results presented above provide a careful and comprehensive analysis of the stability of a TiSe$_2$ monolayer, and establish the $2{\times}2$ PLD as the stable structure in the ground state. Fig.\ \ref{fig:schBS} provides a schematic distillation of the essential physics following from our calculations. Fermi surface nesting is certainly excluded as the cause of this PLD/CDW instability because of the ill defined nesting of the circular/elliptical Fermi surfaces seen both in our results and in experiments  \cite{FermiSurfNest,TiSe2M_CDWNano,TiSe2M_CDWNC}. The absence of any charge density redistribution in a clamped ion $2{\times} 2$ superlattice [cf. Figs. \ref{fig:NormalPhonon}(f) and \ref{fig:NormalPhonon}(g)] indicates that the CDW and PLD are intimately related in the monolayer, similarly to the bulk. It is then clear that the electron-phonon coupling is significant in this system 
because of the large lattice distortions it attains in comparison with similar CDW-prone TMDs \cite{TiSe2B_Th,TMDC_CDWReview}.

Even though the problem is unavoidably interacting and self-consistent, there is a long standing interest in establishing to which extent the CDW is driven here primarily by an electronic instability, or by a strong electron-phonon coupling with negligible influence of electronic correlations (as happens, for example, in metallic TMDs such as NbSe$_2$ \cite{Zhu_CDW}). This is especially important to formulate analytical microscopic models capable of describing the CDW and superconductivity seen in TiSe$_2$ as a function of electron doping, because the presence of strong electronic interactions can affect the pairing instability both quantitative and qualitatively.

It is generally difficult to answer this question from a purely DFT perspective and much less quantify precisely the role of electron-electron interactions because of their approximate treatment in any practical implementation. Nevertheless, combined with the experience and evidence learned from earlier studies of bulk TiSe$_2$, the present results reinforce the view that interactions play a rather consequential role here. One line of evidence arises from the fact highlighted earlier that our calculated distortions and restructured band dispersions in this phase are in good agreement with experimental results, but these properties are seen to 
quantitatively and strongly depend on the type of XC approximation used as mentioned earlier. Similar considerations apply to the stability of the PLD phase at $T{\,=\,}0$ which is not reproduced at the LDA level, for example \cite{Olevano2014}. Since our calculations rely on fully relaxed ions and explicitly converged $k$-sampling, we are confident that this variation reflects directly the different treatment of interaction effects in those implementations of the XC functional. The other line of evidence is related to the spectral reconstruction in the distorted phase, and the reproduction within the GGA of the Mexican hat profile characteristic of the ARPES quasiparticle spectra. Such bandstructure is expected as the self-consistent ground state in the excitonic-insulator scenario, and can be obtained from a mean-field type analysis of the Coulomb interactions between holes at $\Gamma$ and electrons at $M$ based on effective non-interacting bands for the reference (normal) state \cite{Kohn_EI,Money:2009,Wezel:2010}. The reproduction of this at the DFT level is a non-trivial outcome and, in fact, previously seen only in the electronic structure of bulk TiSe$_2$ after the inclusion of $GW$ many-body corrections \cite{Cazzaniga2012} (and apparently in no other electronic system to date). That the monolayer shows such spectral reconstruction without many-body corrections to the GGA bolsters the view that excitonic correlations do play a key role in the CDW transition.

This study underlines that these monolayers are an exciting material platform to study CDW phases in general, have an interesting phase diagram on their own, and will contribute to illuminate the dominant and long-sought mechanism responsible for the CDW instability, both in bulk and monolayer TiSe$_2$.

\section*{acknowledgments}
The work at the National University of Singapore was supported by the Singapore National Research Foundation under the NRF fellowship Award No. 
NRF-NRFF2013-03 (HL), by the Singapore Ministry of Education Academic Research Fund Tier 2 under Grant No. MOE2015-T2-2-059 (VMP), and benefited from the HPC facilities of the NUS Centre for Advanced 2D Materials. WFT is supported by the National Thousand-Young-Talents Program, China.


\begin{thebibliography}{10}
\bibitem{footnote1}
Our calculated critical smearing parameter $\sigma$ strongly depends on the smearing method used in calculations, consistent with Ref. \cite{TiSe2B_ThSm}. It should be noted that while $\sigma$ can be used to access qualitative changes in the phonon spectrum and stability with temperature, its direct relation with the temperature relies on the smearing function used \cite{vasp,Sm_FD,Sm_MethPax,Sm_gauss}.
\bibitem{footnote2}
Note that the phonon spectrum is calculated within the harmonic approximation whereas an estimation of the exact transition temperature ($T_c$) explicitly needs to consider quasi-harmonic effects \cite{phonopy}. The 
$\sigma$ dependence only changes the filling of energy bands without inducing noticeable structural changes \cite{TiSe2B_ThSm}. It will be interesting to estimate $T_c$ in the future through either calculating volume dependence of phonon spectrum \cite{phonopy} or using finite-difference phonon results with large displacements \cite{qha_Antolin,qha_lazar}.
\end{thebibliography}
\end{document}